\documentclass[aps,prc,twocolumn]{revtex4}
\usepackage{graphicx,amssymb}

\begin{document}

\title{Counter-Factual Meaningfulness and the Bell and CHSH Inequalities}
\author{Travis Norsen}
\affiliation{Marlboro College \\ Marlboro, VT  05344 \\ norsen@marlboro.edu}

\date{\today}

\begin{abstract}
We discuss the role of counter-factual meaningfulness (a weaker
cousin of ``counter-factual definiteness'') as a premise in
the derivation of the Bell and CHSH inequalities.  The basic question
motivating the discussion is this:  can the CHSH inequality, unlike
the original Bell inequality, be derived without making a
hidden-variables (or equivalent counter-factual definiteness)
assumption?  We answer, somewhat tentatively, in the negative, and
suggest that an appropriately-modified version of the EPR argument is
needed to rigorously establish that the empirical violation of
Bell-type inequalities can only be blamed on the failure, in nature,
of local causality.
\end{abstract}

\maketitle

\section{Introduction}
\label{intro}

In a recent paper \cite{nonlocchar}, 
I characterized interpreters of Bell's Theorem 
\cite{bell} as falling into two classes:  (1) those who think that
Bell's result (and the associated experimental data) prove that
non-locality is a necessary feature of any empirically viable theory
and hence a feature of nature itself, and (2) those who think that the
results prove merely that non-locality is a necessary feature of any
empirically viable theory of \emph{hidden variables} (HV -- or any theory
containing a \emph{counter-factual definiteness} (CFD) property, or some
other similar anti-orthodox or anti-Copenhagen character).  
\footnote{Please note that here, and throughout this paper, we use the
  term ``locality'' to mean Bell's
  mathematically precise definition of relativistically local
  causality.  See, e.g., \cite[pg 232-248, pg 52-62]{bell} and
  \cite{nonlocchar}.  Likewise, the term ``non-locality'' always
  refers to the failure of Bell's local causality condition.}
In that earlier paper I took it for granted that Bell's derivation of
the (original, Bell) inequality \emph{did} use a HV/CFD assumption,
but argued that a modified (rigorous) version of the EPR argument \cite{epr}
establishes that the necessary HV/CFD principle \emph{follows from
  locality} -- and hence should not be thought of as a separate axiom
behind Bell's inequality, restricting the class of theories to which
the inequality is applicable (and hence likewise restricting the class
of theories which the theorem shows must be non-local in order to agree with
experiment).  In short, I argued that locality \emph{alone} gives rise
to the inequality (but in a two-step dance consisting of the modified
EPR argument followed by the derivation of the Bell inequality), thus
demonstrating the correctness of the interpreters in ``class (1)''
from above.

An anonymous referee for that earlier paper, however, questioned
whether this sophisticated two-step argument was really \emph{needed},
since the CHSH \cite{chsh}
inequality (\emph{unlike} the original Bell inequality,
he suggested) could be derived \emph{without} assuming HV/CFD.  That
is, the referee claimed that the CHSH inequality follows (in one step,
so to speak) from \emph{locality alone} -- and hence its empirical
violation demonstrates already that nonlocality is a fact of nature,
without further analysis or discussion.  \cite{skyrms}

I was initially skeptical of this claim, operating under the
assumption that the difference between the various (generalized-) Bell
inequalities is (roughly) ease of empirical testing, and not anything
about their logical ``inputs''.  After a rather (and, in retrospect,
unfairly and unfortunately) dismissive response to the referee, a
voice in the back of my head urged me to think about this matter
further, which I proceeded to do.  

The first puzzle raised by this further thinking was this:  is it even
true that the (original) Bell inequality has some HV/CFD assumption
built into it?  This, I think, is widely accepted as a truism and
there's no doubt that many of the later attempts to popularize,
explain, and derive a Bell inequality \emph{do} make a HV/CFD assumption
explicitly. (Mermin's various derivations using the idea of
``instruction sets'' \cite{mermin}
are probably the clearest examples here.)  
But where, exactly, does this assumption appear
in Bell's original derivation?  The answer turned out to be more
subtle than I expected.  But finding, eventually, that the answer is
(probably) ``yes'' put me on the path toward eventually identifying a
similar (though not identical, and the difference is subtle and
interesting) sort of HV/CFD assumption that appears also in the
derivation of the CHSH inequality.  

So I am left, in the end, (basically) agreeing with my initial sense,
though also embarrassed at how superficial my prior basis for this
conclusion had been.  I am also left with a much deeper
appreciation for why there has been such long and lingering
disagreement between advocates of (1) and (2) from the first paragraph
above.  For the sense in which HV/CFD appears in the CHSH derivation
is subtle and not obvious, and it is indeed still not crystal clear to
me that this assumption is even really there!  So the final conclusion
in my mind is this:  it's good that the modified EPR argument
(\emph{from} locality \emph{to} HV/CFD) exists to unequivocally settle the
dispute in favor of (1).

But the real point of the current essay is the interesting journey,
not so much the conclusion.  So let us jump in with that.

\section{Bell}
\label{bell}

Let's get notation out of the way first.  Throughout, we will consider
the standard EPR-Bell setup, in which a central source emits
oppositely-directed particles (electrons, say) in the spin-entangled
singlet (total spin zero) state.  (That, at least, is how orthodox QM
describes the relevant state; other candidate theories might give a
more or less or differently detailed account of the particle pair's
state.)  Note in particular that the state is not a spin eigenstate
for either of the individual particles; i.e., orthodox QM (OQM)
attributes no definite spin-component values to either particle
individually prior to measurement.  

After the particles fly apart to some large distance (so that subsequent
spin-component measurements, which take some finite time, can
nevertheless be made with spacelike separation), two experimenters
(Alice and Bob) randomly choose spatial axes ($\hat{a}$ and $\hat{b}$
respectively) along which to measure the spins of their particles.
The outcomes of these measurements are bivalent, and we'll use units
in which the possible outcomes are $A=\pm 1$, $B = \pm 1$.  

Bell's derivation of the inequality gets started as follows:  consider
a theory (not necessarily OQM) according to which a \emph{complete}
description of the state of the particle pair (on some hypersurface
just prior to the measurements, say) is denoted $\lambda$.  Note that
$\lambda$ could be merely the QM wave function, or it could include
more -- or indeed less -- structure attributed to the particles.  The
mere notation ($\lambda$) commits us, really, to nothing.  I stress
this point because there is an unfortunate tendency in the Bell
literature to \emph{call} $\lambda$ ``the hidden variable'' in
contexts in which there is no reason or need to commit to this (i.e.,
to commit to the claim that $\lambda$ attributes additional structure
to the pair's state, compared with OQM's wave function).  This
tendency has no doubt contributed to the confusion about whether, and
how, some HV/CFD type assumption is present in the derivation of the
inequalities.  We will go beyond orthodoxy soon enough, but let's not
fool ourselves into thinking that the mere idea of a complete state
description (or an arbitrary choice of symbol for it) commits us to
anything anti-orthodox.  

Following Bell, let us now assume that the outcomes are
\emph{determined}
\footnote{This is not, however, an arbitrary assumption or axiom.
  Bell argues \emph{for} determinism \emph{from} locality, alluding to
  the EPR argument.  It is unfortunate that this
  first part of Bell's two-part argument for non-locality has been so
  widely forgotten.  I have recently worked to resurrect the argument and
  to clarify its connection to Bell's Theorem -- see, e.g.,
  \cite{nonlocchar} and references therein.  But for purposes of the
  present paper we will simply leave EPR (modified or otherwise) aside
  in order to scrutinize carefully what assumptions go into the
  derivation of the Bell (and, later, CHSH) inequality.}
-- i.e., that there exist functions
\begin{equation}
A(\hat{a},\hat{b},\lambda) = \pm 1
\end{equation}
and
\begin{equation}
B(\hat{a},\hat{b},\lambda) = \pm 1
\end{equation}
which give the outcomes of Alice's and Bob's experiments under
the relevant conditions.  Let us then require \emph{locality} (i.e.,
each outcome is determined without influence from the distant
apparatus orientation), so the relevant functions have the form
\begin{equation}
A(\hat{a},\lambda) = \pm 1
\end{equation}
and
\begin{equation}
B(\hat{b},\lambda) = \pm 1.
\end{equation}
Now, clearly, we are no longer talking about OQM, and for (already)
two distinct reasons.  The first is \emph{determinism}:  Bell assumes
that for a given $\lambda$ and a given $\hat{a}$ and $\hat{b}$, unique
outcomes are determined.  But OQM is not a deterministic theory.
Were we to insist on maintaining allegiance to orthodox quantum
philosophy, we should talk only of \emph{probabilities} for the
various possible outcomes, e.g.,
$P(A|\hat{a},\hat{b},\lambda)$.
(As we will discuss later, the major advance of the CHSH inequality
relative to the original Bell inequality is that it does away with
this assumption of determinism; but more on that later.)

Bell also goes beyond orthodoxy in requiring \emph{locality}.
Orthodox QM simply does not respect Bell's locality condition
(elaborated in \cite{nonlocchar}).  OQM is not a local theory.

To summarize, Bell has us consider a deterministic 
local hidden variable (LHV) theory.  ``Deterministic'' and ``local''
are obvious; the theory being considered is a ``hidden variable''
theory because the complete state description ($\lambda$) cannot
merely be the quantum mechanical wave function (which simply does not
attribute enough structure to the particles to uniquely and locally
determine the outcomes of spin measurements!).

Continuing with the functions $A(\hat{a},\lambda)$ and
$B(\hat{b},\lambda)$, Bell has us consider the expected value of the
product of the outcomes of Alice's and Bob's measurements:
\begin{equation}
E(\hat{a},\hat{b}) = \int d\lambda \; \rho(\lambda)\, A(\hat{a},\lambda)
B(\hat{b},\lambda) 
\label{corrfunc1}
\end{equation}
where $\rho(\lambda)$ is the probability density for the ``singlet
state'' pair preparation procedure to produce the state $\lambda$.  

Bell now imposes the perfect correlation requirement (a special case
of the predictions of OQM):  whenever Alice and Bob measure along the
same axis, they will get opposite outcomes.  Thus
\begin{equation}
A(\hat{b},\lambda) = -B(\hat{b},\lambda)
\label{pcc}
\end{equation}
which allows us to rewrite the correlation function (\ref{corrfunc1}) as
\begin{equation}
E(\hat{a},\hat{b}) = - \int d\lambda \; \rho(\lambda) \, A(\hat{a},\lambda)
A(\hat{b},\lambda). 
\label{corrfunc2}
\end{equation}
This expression contains the first apparent seeds of an additional
(third) anti-orthodox assumption in the derivation.  Taken straight, the
meaning of this expression is something like:
\begin{itemize}
\item
 the expected value for the product of
\begin{enumerate}
\item[(i)] the outcome of Alice's measurement along $\hat{a}$, and
\item[(ii)] the outcome of Alice's measurement along $\hat{b}$.
\end{enumerate}
\end{itemize}
The important point here is \emph{not} that the expression implies the
existence of a \emph{particular} value for the outcome of either
measurement.  That aspect is justified by the already-noted assumption
of determinism.  What's new and crucial here is the (apparent)
requirement that the values $A(\hat{a},\lambda)$ and
$A(\hat{b},\lambda)$ are \emph{simultaneously meaningful}, even though
at most \emph{one} of the measurements (along $\hat{a}$ or $\hat{b}$)
can \emph{actually be performed}.  The whole idea of a meaningful 
correlation
(or more precisely its expectation value) between the outcomes of two
\emph{incompatible} measurements,
surely goes beyond the verificationist/positivist emphasis of the
orthodox framework, in particular Bohr's insistence that 
``the measuring instruments ... serve
to define the conditions under which the phenomena 
appear.''\cite[pg 2]{bell}
 
I find it helpful to think of this orthodox principle as asserting the
``contextuality'' of measurement outcomes:  because (according to the
orthodox view) the outcomes are not pre-encoded in the measured object, but
only arise in the interaction of the  object with the measuring
apparatus, it is meaningless to refer to the outcomes of merely
hypothetical measurements.  ``Unperformed experiments have no
results.''\cite{peres}  Or:  ``No elementary phenomenon is a
phenomenon until it is a registered (observed) phenomenon.''\cite{wheeler}  

So in addition to the determinism and locality assumptions already
noted, we have here a third assumption that might be called
``non-contextuality'' or ``counter-factual meaningfulness'' (CFM -- 
the idea being that it is meaningful to talk about the outcomes of
not-actually-performed measurements).

It is rather difficult, however, to separate this third assumption
from the first-noted assumption (determinism).  The contrast here
would be a theory that is anti-orthodox in (say) the first two ways
(it is deterministic, and it is local) but which is nevertheless
orthodox in the third sense:  according to this 
hypothetical theory, the outcomes
of measurements are brought into being (deterministically and locally)
by an interaction with the measurement apparatus, so while it is
perfectly meaningful to speak of $A(\hat{a},\lambda)$ (in a situation
where a measurement along $\hat{a}$ is actually occuring) or of
$A(\hat{b},\lambda)$ (in a situation where a measurement along
  $\hat{b}$ is actually occuring), it would be impossible to speak
  meaningfully of the product
\begin{equation}
A(\hat{a},\lambda) A(\hat{b},\lambda)
\end{equation}
since the measurement apparatuses needed to make (respectively) the
first and second factors meaningful, are incompatible.  So there can
be no situation in which the product is meaningful.

Actually, the last few paragraphs have been deliberately (but
justifiably) misleading.  This third anti-orthodox assumption
(CFM or non-contextuality) is actually
\emph{not} implied by the mere writing of Equation (\ref{corrfunc2})
-- a point that is clear if we just remember where Equation
(\ref{corrfunc2}) came from.  Recall that the apparently problematic
expression came from the combination of the unproblematic definition
of the correlation function -- Equation (\ref{corrfunc1}) -- with the
perfect correlation condition -- Equation (\ref{pcc}).  Remembering this
allows us to give meaning to the correlation as expressed in Equation
(\ref{corrfunc2}) \emph{without} making any CFM
assumption, simply by reading the factors as
\begin{enumerate}
\item[(i)] $A(\hat{a},\lambda) \, =$ the value the theory predicts for
  $A$ when Alice measures along $\hat{a}$ (with the pair in state
  $\lambda$), and
\item[(ii)] $A(\hat{b},\lambda) \, =$ the \emph{opposite} of the 
  value the theory predicts for $B$ when Bob measures along $\hat{b}$
  (with the pair in state $\lambda$).
\end{enumerate}
And since the two relevant measurements here (Alice's along $\hat{a}$
  and Bob's along $\hat{b}$) are perfectly compatible, we need not
  assume the meaningfulness of any measurement outcomes which are not
  (and indeed cannot be) actually performed.

By parsing it this way, we see that this 
third sort of anti-orthodoxy is not actually present in
Equation (\ref{corrfunc2}).  But our reason for going into this is
that this anti-orthodox sort of CFM \emph{is} assumed in the
\emph{next} step in Bell's derivation, and in a way that cannot be
eliminated by any re-parsing of the problematic mathematical
expressions.  Let us see how this comes about by following through
with Bell's derivation.  Consider the difference in correlation
functions, expressed as in Equation (\ref{corrfunc2}), 
for two different pairs of angles:
\begin{eqnarray}
&& E(\hat{a},\hat{b}) - E(\hat{a},\hat{c}) \\
&& \; =  \int  d\lambda \; \rho(\lambda) \, \left[ A(\hat{a},\lambda)
  A(\hat{c},\lambda) - A(\hat{a},\lambda) A(\hat{b},\lambda)
\right]. \nonumber 
\end{eqnarray}
So far so good.  But now we insert unity, into the first term in the
square brackets, in the form
\begin{equation}
1 = A(\hat{b},\lambda) A(\hat{b},\lambda)
\end{equation}
(justified by the idea that $A(\hat{b},\lambda)= \pm 1$ so that,
either way, its square is one).  This leaves us with
\begin{eqnarray}
&& E(\hat{a},\hat{b}) - E(\hat{a},\hat{c}) \nonumber \\
&& = \int d\lambda \; \rho(\lambda) \, \left[
A(\hat{a},\lambda) A(\hat{b},\lambda) A(\hat{b},\lambda)
A(\hat{c},\lambda) \right.  \nonumber \\
& & \qquad \qquad \qquad \qquad
 - \left. A(\hat{a},\lambda) A(\hat{b},\lambda)
\right] \; \; 
\label{bell4}
\\
&& = \int d\lambda \; \rho(\lambda) \, A(\hat{a},\lambda) A(\hat{b},\lambda) 
\left[ A(\hat{b},\lambda) A(\hat{c},\lambda) - 1 \right]
\end{eqnarray}
From here it is straightforward to get Bell's inequality.  Taking
absolute values on both sides, using the fact that $|A| \le 1$, and
identifying 
\begin{equation}
-\int d\lambda \; \rho(\lambda) \, A(\hat{b},\lambda) A(\hat{c},\lambda) =
 E(\hat{b},\hat{c})
\end{equation}
gives Bell's inequality:
\begin{equation}
\left| E(\hat{a},\hat{b}) - E(\hat{a},\hat{c}) \right| \le 1 +
E(\hat{b},\hat{c}).
\end{equation}

The reader has probably already noticed the step to which we wish to
call attention.  One of the terms in Equation (\ref{bell4}) reads
\begin{equation}
\int d\lambda \; \rho(\lambda) \, A(\hat{a},\lambda) A(\hat{b},\lambda)
A(\hat{b},\lambda) A(\hat{c},\lambda)
\label{fourfactors}
\end{equation}
which, in words, evidently means
\begin{itemize}
\item
 the expected value for the product of
\begin{enumerate}
\item[(i)] the outcome of Alice's measurement along $\hat{a}$,
\item[(ii)] the outcome of Alice's measurement along $\hat{b}$
  (squared), and
\item[(iii)] the outcome of Alice's measurement along $\hat{c}$.
\end{enumerate}
\end{itemize}
Now the same kind of analysis we gave before will go through, but this
time without the escape hatch that saved us before -- namely, the
re-parsing of one of the two apparently-incompatible factors in terms
of another, actually-compatible measurement (namely Bob's).   For even
supposing we parse one of the group \{ (i), (ii), and (iii) \} in terms of
a measurement made by Bob, we are here left inevitably with two
incompatible measurement contexts (e.g., $\hat{a}$ and $\hat{b}$) for
Alice.  And so, from the point of view of any theory that respects this third
orthodox principle, such an expression as Equation (\ref{fourfactors})
is simply \emph{meaningless} -- and therefore anything that comes
after its appearance (such as Bell's inequality) is equally
meaningless.  And hence Bell's inequality simply would not apply to
the kind of theory we raised the possibility of before:  one which is
deterministic and local, but which accepts the orthodox principle of
contextuality (i.e., which denies CFM).

Summing up, it appears that Bell's derivation of the inequality is
premised on three distinct assumptions:
\begin{itemize}
\item determinism
\item locality
\item CFM (or non-contextuality)
\end{itemize}
any of which might, in principle, be rejected in the face of empirical
data contradicting the inequality.  
\footnote{Remember, however, that this is the case only if one leaves
  aside the modified EPR argument, which establishes that
  non-contextual, deterministic hidden variables follow from locality
  plus the empirical fact of perfect anti-correlation.  So, really,
  there is no choice.  The empirical violations of Bell's inequality
  can only be blamed on the locality assumption.  But remember that
  our goal here is to explore the situation without the modified EPR
  argument, and in particular to see how the CHSH inequality fares in
  this regard relative to Bell's.}

One point should be addressed here before finally moving on to
consider the CHSH inequality.  Someone skeptical of the previous
discussion could raise the following objection:  surely Alice needn't
\emph{actually} set up a measurement apparatus oriented along
$\hat{b}$ in order to justify the statement that $A(\hat{b},\lambda)
A(\hat{b},\lambda) = 1$.  Even a merely \emph{imagined} measurement
along $\hat{b}$ (even one which \emph{conflicts} with an
\emph{actually} performed measurement along, say, $\hat{a}$) must obey
$A(\hat{b},\lambda) = \pm 1$, and hence $A(\hat{b},\lambda)^2 = 1$.
Can't we then parse the allegedly-objectionable factors in Equation 
(\ref{fourfactors}) in terms of such an imaginary measurement, thus
removing the need for any CFM-type assumption?

The problem with (i.e., the answer to) this objection is that
``imagination'' provides far too much leeway.  For example, couldn't
we imagine that $A(\hat{b},\lambda) = +1$ ... and then imagine that
$A(\hat{b},\lambda) = -1$ ... so the product is $-1$ rather than the
required $+1$?  Or, for that matter, couldn't we imagine
$A(\hat{b},\lambda) = 0$ or $17 \pi / \sqrt{2}$ or any other value
whose square is not $+1$?  Of course, the objector will want to say
that the imaginations must be constrained by the (local deterministic)
theory, which must (by prior assumptions) attribute either the value
$+1$ or the value $-1$ to Alice's (real or imaginary) measurement
along $\hat{b}$.  But this just brings out the fundamental problem
with the objection.  Such a theory need only attribute this value to
Alice's measurement for an experimental context in which a measurement
along $\hat{b}$ actually occurs.  There are no constraints whatever on
what value (if any) such a theory must attribute to a measurement
along the $\hat{b}$ direction \emph{in an experimental context in
  which Alice is actually measuring along the $\hat{a}$ direction}.
Indeed, in principle, for a contextual theory, there \emph{can be} no
such attributed value, for the very idea being considered (the outcome
of a measurement that is incompatible with another actually performed
measurement) is simply meaningless.

And so the original claim -- that Bell's derivation of the inequality
requires not only locality and determinism, but also a CFM-type
assumption -- stands.

\section{CHSH}

The main difference between the Bell inequality and the CHSH
inequality is that the latter does not require determinism.  Instead
of beginning with an assumption that the outcomes $A$ and $B$ are
\emph{fixed} once the experimental context ($\hat{a}$, $\hat{b}$) and
particle-pair state ($\lambda$) are specified, CHSH require only that
a theory specify the probabilities for the various possible outcomes.
Thus, instead of functions $A(\hat{a},\lambda)$ and
$B(\hat{b},\lambda)$, we begin with two probability functions:
\begin{equation}
P(A|\hat{a},\lambda)
\end{equation}
and
\begin{equation}
P(B|\hat{b},\lambda).
\end{equation}
Note that we have already imposed the locality condition, whereby the
probability (assigned by the theory in question to the various
possible outcomes) depends only on facts which are locally accessible
to the experiment in question.  Thus, for example, 
the probability for different
outcomes $A$ depends only on $\hat{a}$ and $\lambda$ -- not on the
distant setting $\hat{b}$ or the distant outcome $B$.  (Note that it
is only because the state description $\lambda$ is assumed to be a
\emph{complete} description, that the non-dependence of the probability
of $A$ on the distant outcome $B$ follows from local causality.)  

For such a (local, but not necessarily deterministic) theory, the
expectation value of the product of the two outcomes will be given by:
\begin{equation}
E(\hat{a},\hat{b}) = \int d\lambda \; \rho(\lambda) \, \sum_{A,B} A \,
B \, P(A|\hat{a},\lambda) \, P(B|\hat{b},\lambda).
\end{equation}
Since $A=\pm 1$ and $B=\pm 1$, this can be simplified to
\begin{equation}
E(\hat{a},\hat{b}) = \int d\lambda \; \rho(\lambda) \,
\bar{A}(\hat{a},\lambda) \, \bar{B}(\hat{b},\lambda)
\label{chshcorr}
\end{equation}
where
\begin{eqnarray}
\bar{A}(\hat{a},\lambda) &=& \sum_A \; A \; P(A|\hat{a},\lambda) \nonumber \\
&=& P(A=+1 | \hat{a},\lambda) - P(A=-1 | \hat{a},\lambda) 
\end{eqnarray}
is the theory's prediction for the average value of Alice's experiment 
(with the apparatus set along $\hat{a}$ and with particles in the
state $\lambda$).  (And similarly for $\bar{B}$.)

Thus parsing Equation (\ref{chshcorr}):  the expected value of the
product of Alice's and Bob's experiments is given by a weighted
average (over all the pair states that could be produced by the
preparation procedure) of the product of
\begin{enumerate}
\item[(i)] the average value for Alice's measurement, and
\item[(ii)] the average value for Bob's measurement
\end{enumerate}
where the averages here are averages over the \emph{possible} outcomes
allowed by the theory when
the experimental context $\hat{a}$ and $\hat{b}$ is realized.

Continuing with the derivation, let us consider, as before, the
difference in the correlation function for two different pairs of
angles:
\begin{eqnarray}
&&E(\hat{a},\hat{b}) - E(\hat{a},\hat{b}') = \\
&& \; \int d\lambda \; \rho(\lambda) \,
\left[ \bar{A}(\hat{a},\lambda) \bar{B}(\hat{b},\lambda) 
      - \bar{A}(\hat{a},\lambda) \bar{B}(\hat{b}',\lambda) \right]
    \nonumber 
\end{eqnarray}
where $\hat{b}$ and $\hat{b}'$ refer to two distinct settings of Bob's
apparatus (as will, likewise, $\hat{a}$ and $\hat{a}'$ refer shortly
to two distinct settings of Alice's apparatus).  So far so good.  But
now, to continue with the derivation, we need to add zero inside the
integrand in the clever form
\begin{eqnarray}
0 & = & \pm \bar{A}(\hat{a},\lambda) \bar{A}(\hat{a}',\lambda)
\bar{B}(\hat{b},\lambda) \bar{B}(\hat{b}',\lambda) \nonumber \\
&&\; \; \mp \bar{A}(\hat{a},\lambda) \bar{A}(\hat{a}',\lambda)
\bar{B}(\hat{b},\lambda) \bar{B}(\hat{b}',\lambda).
\end{eqnarray}
Rearranging and factoring \cite{clover} inside the integrand then gives
\begin{eqnarray}
&&E(\hat{a},\hat{b}) - E(\hat{a},\hat{b}') = \nonumber 
\\
&& \; \int d\lambda \; \rho(\lambda) \,
 \left[   \bar{A}(\hat{a},\lambda) \bar{B}(\hat{b},\lambda) \left( 1 \pm
\bar{A}(\hat{a}',\lambda) \bar{B}(\hat{b}',\lambda) \right) \right. \nonumber
\\
&& \; \; \; \; \; \; \;  
- \left. \bar{A}(\hat{a},\lambda) \bar{B}(\hat{b}',\lambda) \left( 1 \pm
\bar{A}(\hat{a}',\lambda) \bar{B}(\hat{b},\lambda) \right)  \right] 
\label{chshcorrdiff}
\end{eqnarray}
which is easily reduced, by taking absolute values on both sides, to
\begin{equation}
\left| E(\hat{a},\hat{b}) - E(\hat{a},\hat{b}') \right| +
\left| E(\hat{a}',\hat{b}') + E(\hat{a}',\hat{b}) \right| \le 2
\end{equation}
which is the CHSH inequality.

The reader has probably already noticed the step to which we wish to
draw attention.  The two terms (which add to zero) that appear for the
first time in Equation (\ref{chshcorrdiff}) have the form
\begin{equation}
\int d\lambda \; \rho(\lambda) \,  \bar{A}(\hat{a},\lambda) \,
\bar{A}(\hat{a}',\lambda) \, \bar{B}(\hat{b},\lambda) \,
\bar{B}(\hat{b}',\lambda) 
\label{chshfour}
\end{equation}
which, in words, is evidently
\begin{itemize}
\item
 the expected value for the product of
\begin{enumerate}
\item[(i)] the average value for the outcome of 
             Alice's measurement along $\hat{a}$,
\item[(ii)] the average value for the outcome of Alice's measurement 
             along $\hat{a}'$,
\item[(iii)] the average value for the outcome of Bob's measurement 
             along $\hat{b}$, and
\item[(iv)]  the average value for the outcome of Bob's measurement
             along $\hat{b}'$.
\end{enumerate}
\end{itemize}
which, like the similar terms discussed in the previous section, would
only seem to be \emph{meaningful} for a theory of non-contextual
hidden variables.  That is, for a theory according to which the
experimental outcomes are brought into existence by some kind of
interaction between the measuring apparatus and the measured object,
it would be meaningless to talk simultaneously about the outcomes of
incompatible measurements (such as Alice's measurements along both
$\hat{a}$ and $\hat{a}'$).

Of course, one might object, we are not here forced, by the algebra,
to talk about \emph{specific} outcomes for these pairs of incompatible
measurements.  Instead, we only need simultaneous talk about the
\emph{averages} of those pairs of incompatible measurements.  This
objection is correct as far as it goes:  the assumption here is indeed
weaker than the full ``counter-factual \emph{definiteness}'' which is
used in the derivation of the original Bell inequality.  But there is
still here, in the CHSH derivation, an assumption of the weaker
condition we have called CFM:  counter-factual meaningfulness.  This
parallels closely the discussion in the previous section, so we need
not elaborate in great detail.  Suffice it to point out that locality
alone (which is the only \emph{explicit} assumption noted so far in
the derivation of the CHSH inequality) does not (in any obvious way)
warrant a claim that
the average value for a measurement by Alice along $\hat{a}'$ should
be the same under two scenarios:  first, the measurement along $\hat{a}'$ is
actually performed, and second, the measurement along $\hat{a}'$ is
performed while the apparatus is oriented along $\hat{a}$.  Indeed, it
is difficult to understand how a theory could yield up \emph{any}
determinate value for the average under the second set of conditions
\emph{since those conditions are in principle unrealizable -- they
  involve a contradiction, because the apparatus cannot be
  simultaneously aligned along both $\hat{a}$ and $\hat{a}'$}.

Of course, there is no such problem in a theory of 
non-contextual hidden variables -- i.e., a theory in which the outcomes
of experiments (or the probabilities for various possible outcomes)
are pre-encoded in the state of the object alone, with the role of the
experimental apparatus being simply to \emph{reveal} those
pre-existing values (or to reveal one of the possible values according
to the pre-existing probability distribution that is, so to speak,
encoded in the state of the ``measured'' object).  In such a theory,
we can understand the problematic expressions such as Equation
(\ref{chshfour}) as referring, not to (averages of) outcomes of
actually-performed experiments, but to the hidden variables themselves
-- to those features of the state of the object which determine what
the averages will be \emph{if} a measurement is performed (but with no
implication that such a measurement need \emph{actually be} performed 
in order to give meaning to the expression).  

We thus conclude that, in addition to the assumption of locality
(which nobody denies is present), the derivation of the CHSH inequality
requires also a second assumption:  not the ``counter-factual
definiteness'' (CFD) that is needed to derive the Bell inequality, that is
true, but a weaker condition of counter-factual \emph{meaningfulness}
(CFM, which can be roughly thought of as ``CFD minus determinism'').
CFM is the assumption that underwrites simultaneous talk about (the
averages of) outcomes of incompatible experiments, and is practically
equivalent to the assumption of non-contextuality (which is violated
by both orthodox QM and contextual hidden variable theories such
as Bohmian Mechanics, both of which are, however, non-local theories).  
Thus, the only familiar examples of theories satisfying the two
conditions needed to derive the CHSH inequality would be \emph{local,
non-contextual hidden variable theories}.  We therefore conclude that
it is misleading (or flat wrong) to assert that the CHSH inequality is
an example of a Bell inequality ``without hidden
variables.'' \cite{eberhard}   It's true that one need not explicitly assume
hidden-variables in order to arrive at the inequality; but the CFM
assumption one \emph{does} need is, for all practical purposes,
equivalent.

\section{Discussion}

We have identified three assumptions 
\begin{enumerate}
\item determinism
\item locality
\item counter-factual meaningfulness (CFM)
\end{enumerate}
which function as premises in the derivation of the Bell inequality
(which requires all three premises) and the CHSH inequality (which
requires only the second and third).  

The status of the CHSH inequality relative to the Bell inequality has
been obscured, in previous literature, by the packaging of our
premises 1 and 3 into ``counter-factual definiteness'' (CFD) -- the
idea that there is some uniquely determined outcome to
not-actually-performed measurements.  This packaging has apparently
led some people to the erroneous conclusion that, since the full CFD
property is not assumed in the derivation of the CHSH inequality, it
is based on no assumption other than locality.  But this is false.
One does still need the (weaker) CFM principle (and locality) to
arrive at CHSH.  And so, in principle \footnote{...and again leaving
  aside EPR...} a violation of the CHSH inequality could be blamed
\emph{either} on a failure of locality \emph{or} a failure of
counter-factual meaningfulness (i.e., a failure of non-contextuality).  

In another nice paper \cite{peres}
that argues for a Bell-type inequality making
``no mention ... of `hidden variables' or similar superstitions'',
Asher Peres characterizes the logical status of the inequality as
follows:  ``Let us assume that the outcome of an experiment performed
on one of the systems is independent of the choice of the experiment
performed on the other.  Now, let us try to imagine the results of
alternative measurements, which could have been performed on the same
systems \emph{instead} of the actual measurements.  Then there is no
way of contriving these hypothetical results so that they will satisfy
all the quantum correlations with the results of the actual
measurements.''  Peres also discusses how the derivation ``involves a
comparison of the results of experiments which were actually
performed, with those of hypothetical experiments which could have
been performed but were not'' and points out that ``it is
\emph{impossible to imagine} the latter results in a way compatible
with (a) the results of the actually performed experiments, (b) long
range separability of results of individual measurements, and (c) [the
empirical predictions of] quantum mechanics.''

What then should we infer from the fact that the inequalities are
empirically violated?  Peres tells us that ``[t]here are two possible
attitudes in the face of these results.  One is to say that it is
illegitimate to speculate about unperformed experiments.  In brief
`Thou shalt not think.'  ....  Alternatively, for those who cannot
refrain from thinking, we can abandon the assumption that the results
of measurements by A[lice] are independent of what is being done by
B[ob].''  These statements nicely summarize the conclusions of the
current essay.  Unlike many other authors, Peres correctly
characterizes the additional assumption (beyond Locality) needed to
derive the CHSH inequality as weaker than ``counter-factual
definiteness'' -- it is, rather, the assumption that it is meaningful
to speculate \emph{at all} about unperformed experiments.  
It's not just that one shouldn't think of them as having
definite outcomes; one cannot even think of them as having a
well-defined average outcome.  In short, following Peres, one ``shalt
not think'' about the results of un-performed (and/or un-performable)
experiments \emph{at all}.

A theory which maintains allegiance to this orthodox principle (``Thou
shalt not think'' about un-performed experiments) could make
predictions consistent with the inequalities (and hence, now, with
experiment) even if it were local.  Or so it might seem.  But before
spending any time trying to concoct such a theory (which, if found,
would be an explicit counterexample refuting the claims of those -- from
``class (1)'' mentioned in the introduction -- who think that the
empirical violations of Bell's inequalities proves that nature is non-local)
one would do well to remember the existence of the modified EPR
argument (as detailed in \cite{nonlocchar}).  For this (unfortunately
neglected, if not completely forgotten) argument proves that the
various other principles needed to arrive at a Bell-type inequality can
all be \emph{derived} from the assumption of locality -- that locality
\emph{requires} a local, non-contextual hidden-variables theory (to
explain the fact of perfect anti-correlation when Alice and Bob measure
along the same axis).  So when we bring this modified EPR argument
back in, we see that, after all, there is no choice but to blame the
empirical violations of Bell-type inequalities on the failure of the
locality assumption.  We cannot, after all, blame the Bell-violating
data on a failure of CFM or any other principle going beyond
orthodoxy (other than locality).  

So our real conclusion is this:  
it is nice that we \emph{don't} need to leave aside the
modified EPR argument (as we have done throughout the current paper).
Without that other argument (the first half of Bell's own two-part
argument for non-locality) we might accidently \emph{fool ourselves}
into thinking that the empirical data can be dealt with by rejecting
something other than locality.  But it can't -- a realization
which leaves us in a better position 
to appreciate Bell's statement that ``For me then this is
the real problem with quantum theory:  the apparently essential
conflict between any sharp formulation and fundamental relativity.
That is to say, we have an apparent incompatibility, at the deepest
level, between the two fundamental pillars of contemporary theory...''
\cite[pg 172]{bell}

\section{Confession}

Aside from a few parenthetical qualifications snuck into the abstract
and introduction, I have tried to present the arguments here as
forcefully and univocally as possible.  But I feel I should confess to
being, myself, not at all convinced that these arguments are 
correct.  The main thesis I've argued for is that certain matematical
expressions, which necessarily appear in the middle-stages of the
derivation of generalized Bell inequalities, are simply
\emph{meaningless} from the point of view of \emph{contextual}
theories (i.e., theories rejecting CFM), since these expressions
contain a kind of implicit reference to incompatible pairs of
measurements.  

But is there really a new and separate CFM assumption here?  For
example, in the original Bell derivation, once we allow the existence
of \emph{functions} such as $A(\hat{a},\lambda)$, haven't we already
tacitly allowed that these functions are well-defined?  That is, can't
we simply regard $A(\hat{a},\lambda)$ and $A(\hat{a}',\lambda)$ as
``just the same functions ... with different argument'' (as Bell puts
it in answering a related objection in his paper ``Locality in quantum
mechanics: reply to critics'' \cite{bell})?  That is, don't the kind
of hidden variables that are \emph{already} on the table, based on the
determinism and locality assumptions, support simultaneous talk of
$A(\hat{a},\lambda)$ and $A(\hat{a}',\lambda)$ -- as simply the values
that the theory yields for outcomes along two possible (but not
necessarily actual, and by no means ``actually simultaneous'')
measurements?  

And then can't one argue in parallel for the average values that are
used in the context of the CHSH derivation, thus concluding that,
indeed, the CHSH inequality follows (in one step) from the locality
assumption alone (just as the previously-mentioned anonymous referee 
claimed)?

The arguments presented in the earlier sections of this paper have the
following basic structure:  certain mathematical expressions appearing
in the intermediate stages of the algebraic derivation of Bell-type
inequalities, can be ``translated back'' into prose descriptions of
certain correlation functions -- i.e, expectation values for products
of certain sets of measurement outcomes.  See, for example, Equation 
(\ref{fourfactors}) and the subsequent prose translation.  This ``back
translation'' is supposed to have been justified by the fact that it
is merely doing, in reverse, what we did originally to write down a 
mathematical expression -- Equation (\ref{corrfunc1}) -- for ``the
expected value of the product of Alice's measurement along $\hat{a}$
and Bob's measurement along $\hat{b}$''.  And then, according to our
earlier argument, since the ``back translated'' prose statement is 
operationally meaningless (because it refers to the product of 
outcomes of incompatible measurements), so is the corresponding
mathematical expression.

But who says we need to -- or, indeed, are entitled to -- make this
``back translation''?  By the time we get to the relevant stage in the
algebraic derivation, there is no question about the meaningfulness of
the individual factors in (for example) Equation (\ref{fourfactors}).
Each is simply the outcome that the theory in question predicts for
the measurement in question.  And if each of those is individually
meaningful, how in the world can there be any problem in multiplying
them together (and then averaging the result over the possible states
$\lambda$ that might be, according to the theory, produced by the
preparation procedure)?  Sure, if we translate the math back into
prose in a certain way, we get something that is operationally
meaningless.  But we could just as easily -- and, I would argue, more
faithfully -- translate the mathematical expression into prose this
way:  Equation (\ref{fourfactors}) represents
\begin{itemize}
\item
 the average (over possible $\lambda$s) of the product of
\begin{enumerate}
\item[(i)] the value that the theory predicts for the outcome of a
  measurement by Alice along $\hat{a}$

\item[(ii)] the value that the theory predicts for the outcome of a
  measurement by Alice along $\hat{b}$ (squared), and

\item[(iii)] the value that the theory predicts for the outcome of a
  measurement by Alice along $\hat{c}$.
\end{enumerate}
\end{itemize}
And this way there is no reference to the actual outcomes of
actually-performed measurements that are incompatible, no tacit
assumption of CFM.  We are simply talking about what some theory says
will happen in various circumstances, and mathematically manipulating 
values that are perfectly well defined.  There seems to then be no
problem -- no mysterious third anti-orthodox assumption -- whatsoever.

I think the fundamental point here is one that has already been
emphasized, from a slightly different point of view, in
\cite{nonlocchar}.  Confusion arises when one forgets (in the context
of analyzing Bell's theorem) that one is not talking directly about
measurement outcomes, but about \emph{theories} and their
predictions.  (This confusion is probably prevalent because of the
influence of philosophical positivism on the quantum founding fathers and
their followers.)  In the context of the present paper, it seems that as
long as one keeps this in mind -- as long as one resists the
temptation to require a direct operationalist interpretation for every
intermediate stage in the algebraic derivation -- it emerges that
there is no distinct ``CFM'' assumption in the derivation of the Bell
or CHSH inequalities.  And hence it emerges that the CHSH inequality
in particular follows from locality \emph{alone}, such that its
empirical violation can only be blamed on the non-local character of
nature.  

Nevertheless (a meta-confession) I am not 100\% certain that my own
objection to my own arguments is correct.  So, at least for the time
being, I am relieved that the approach taken in \cite{nonlocchar}
exists -- that is, the approach of using a modified EPR argument as 
an ``end run'' around the question of CFM.  The associated two-part
argument for nonlocality still seems (to me) to be the most
straightforward, least subtle, and most airtight proof that locality
(and, in this context, nothing else) has been empirically refuted.

\section*{Acknowledgements:}

Thanks to the anonymous referee of the earlier paper for stimulating
my thinking on this question (and, I hope, for forgiving my earlier
dismissiveness).

\end{document}